\documentclass[%
preprint,
 amsmath,amssymb,
floatfix,
titlepage
]{revtex4-2}
\usepackage[section]{placeins}
\usepackage{float}
\usepackage{empheq}
\usepackage{hyperref}
\usepackage{cleveref}
\usepackage{tikz}
\usepackage{cancel}
\usepackage{caption}
 \usepackage{subcaption}
\usepackage{sidecap}
\usepackage[colorinlistoftodos,prependcaption]{todonotes}
\newcommand{\be}{\begin{equation}}
\newcommand{\ee}{\end{equation}}

\newcommand{\bea}{\begin{eqnarray}}
\newcommand{\eea}{\end{eqnarray}}

\usepackage{slashbox}
\begin{document}
\title{Ring structural transitions in strongly coupled dusty plasmas}
\author{Vikram Dharodi}
\email{vsd0005@auburn.edu}
\author{Evdokiya Kostadinova}
\email{egk003@auburn.edu}
\affiliation{Department of Physics, Auburn University, Auburn, Alabama 32849, USA}
\date{\today}

\begin{abstract}

This paper presents a numerical study of ring structural transitions in strongly coupled dusty plasma confined in a ring-shaped (quartic) potential well with a central barrier, whose axis of symmetry is parallel to the gravitational attraction. It is observed that increasing the amplitude of the potential leads to a transition from a ring monolayer structure (rings of different diameters nested within the same plane) to a cylindrical shell structure (rings of similar diameter aligned in parallel planes). In the cylindrical shell state, the rings alignment in the vertical plane exhibits hexagonal symmetry. The ring transition is reversible, but exhibits hysteresis in the initial and final particle positions. As the critical conditions for the transitions are approached, the transitional structure states exhibit zigzag instabilities or asymmetries on the ring alignment. Furthermore, for a fixed amplitude of the quartic potential that results in a cylinder-shaped shell structure, we show that additional rings in the cylindrical shell structure can be formed by decreasing the curvature of the parabolic potential well, whose axis of symmetry is perpendicular to the gravitational force, increasing the number density, and lowering the screening parameter. Finally, we discuss the application of these findings to dusty plasma experiments with ring electrodes and weak magnetic fields. 

\end{abstract}

\maketitle

%


\section{Introduction}\label{Introduction}

Complex (or dusty) plasmas are collections of nano-sized or micron-sized solid particles suspended in plasma environment. Typically, the dust grains acquire negative charge and interact via the Yukawa (shielded Coulomb) potential. Depending on the coupling strength \cite{ichimaru1982strongly}, the dusty plasma structure can be treated as a fluid~\cite{rao1990dust,ma1997fluid}, a visco-elastic fluid~\cite{kaw1998low,singh2014visco,diaw2015generalized,dharodi2020rotating}, or a crystal \cite{ikezi1986coulomb,thomas1994plasma,chu1994direct,hayashi1994observation}. Variations of the coupling strength lead to phase or structure transitions ~\cite{bin2003structure,melzer2012phase,maity2019molecular} and control over the growth of instabilities (gravity driven~\cite{dharodi2021numericalI,dharodi2021numericalII} and shear driven~\cite{dharodi2022kelvin}), turbulence \cite{gupta2014kolmogorov,zhdanov2015wave,tiwari2015turbulence,kostadinova2021fractional}, and wave propagation~\cite{merlino201425,dharodi2016sub,kostadinova2018transport}), etc. In Earth-based experiments, due to the macroscopic size, the dust particles normally levitate close to the lower electrode, in the plasma sheath, where the gravitational force is balanced by the sheath electric force. In addition, the repulsive Yukawa interaction (the expansion) is commonly balanced by applying an external magnetic field \cite{schwabe2011pattern,thomas2015observations,thomas2012magnetized,konopka2000rigid,choudhary2020three} or by externally applied radial confinement potential \cite{chaubey2022preservation}, for example, due to a disc cutout or a ring placed on the lower electrode of a capacitively-coupled RF reference cell. Thus, the engineering of the vacuum chamber electrodes can be used to shape the confinement potentials in the plasma, which allows for the exploration of a wide range of structural and dynamical phenomena. For example, one-dimensional transverse optical modes have been investigated experimentally using a horizontally aligned (perpendicular to gravity) dust chains confined in a harmonic potential created by a linear groove in the lower electrode ~\cite{liu2003transverse}. The two-dimensional zigzag transitions have been studied in dusty plasmas confined by a biharmonic potential well created by a rectangular depression between four conducting bars placed on the RF powered electrode~\cite{sheridan2010dimensional}. It has been shown that the properties of dust cluster rotation in a non-magnetized dusty plasma is highly dependent on the characteristics of the parabolic radial confinement potential~\cite{huang2013cluster,hartmann2019self}. Numerical simulations of dusty plasma crystals confined in this type of potential have shown a transition from fully hexagonal structure to a structure with hexagonal lattice interior surrounded by concentric rings ~\cite{qiao2007structure}. The formation of ring structure in dusty plasmas is of particular interest to the present study. A ring-shaped quadratic potential well has been used to numerically demonstrate the formation of complete and incomplete dust rings in the horizontal (perpendicular to gravity) plane~\cite{schweigert1996properties,sheridan2009dusty}. Moreover, longitudinal and transverse dispersion relationships have been  experimentally observed for this potential type~\cite{sheridan2016dusty}. A ring-shaped potential was formed using a circular grooved electrode with a center post and used to study a rotating ring of dust particles in a non-magnetized plasma~\cite{theisen2020rotating}. A rotating dust ring has also been observed within the ring-shaped asymmetric potential well created by asymmetric sawteeth of gears on the lower electrode~\cite{he2020experimental}. Finally, it has been shown that in the presence of a weak magnetic field (about $150~G$), the confinement potential of a ring placed on the lower electrode is modified, leading to the formation of a rotating ring dust structure \cite{konopka2000rigid}. 

 Motivated by these experimental observations, here we use molecular dynamics (MD) simulations to explore ring structural transitions for dust particles confined by a ring-shaped (quartic) potential well with a central barrier. Specifically, we investigate the critical conditions leading to a transition from a ring monolayer structure (rings of different diameters nested within the same plane) to a cylindrical shell structure (rings of similar diameter aligned in parallel planes). It is observed that the structural transition is governed by a competition between the strength of the Yukawa interaction potential and the properties of the external confinement potential. Thus, we conjecture that these ring transitions can be used to investigate dust particle charge in experiments where the shape of the double-well potential can be varied (for example, through changing the power on a ring electrode and the strength of an external magnetic field). In the present simulation, two types of external potentials have been employed: a ring-shaped quatric potential well providing the horizontal confinement and a parabolic potential well providing the vertical confinement. The axis of symmetry of the quartic potential is parallel to the gravitational force, whereas the parabolic potential has perpendicular to it. Before proceeding, here, it is also important to keep in mind that if the confinement/direction/plane is horizontal, it is perpendicular to gravity; if it is vertical, it is parallel to gravity. 
 
 For a fixed number of dust particles, the transition from a circular monolayer to a cylindrical shell structure is observed as the amplitude of the quatric potential is gradually increased. The transition is reversible and occurs through several intermediate transitional states. In these intermediate states, since the number of dust particles is either slightly higher or slightly lower than the requirement of ring formation number, these transition states exhibit a zigzag instability \cite{melzer2006zigzag} or formation of uneven rings in both the horizontal and vertical directions. We observe that for an appropriate number of particles in the cylindrical shell phase (the number needed to form perfect rings for a given width of the annular potential well),  the ring alignment in the vertical plane exhibits a hexagonal symmetry. For a fixed number of particles and a fixed amplitude of the quartic potential that supports a cylinder-shaped shell structure, decreasing the curvature of the parabolic potential (causes for the vertical confinement) results in the formation of additional rings in the cylindrical shell structure. This process occurs via several intermediary transition structures that exhibit irregularities only in the vertical direction. Similar observations have been made for simulations where the screening parameter is decreased and/or the dust number density is increased.

 This paper is organized as follows. Section \ref{model_methods} presents the numerical scheme with a brief description of the external forces and potentials involved in our study. Section \ref{Num_Result_Dis} is devoted to the numerical investigation of the structural transition phenomena followed by a detailed discussion of the obtained results. Finally, in Sec. \ref{Conclusions}, we provide a summary of our results  and discuss applications of these findings. The $CGS$ system of units is used in everything that follows. 

\section{Model and methodology}\label{model_methods}

We consider a dusty plasma that includes $N$ dust particles which interact through a Yukawa potential energy ${\mathcal{U}}^{ykw}$. Each particle has the same negative charge $Q$ (in special circumstances like secondary electron emission dust particles become positively charged~\cite{shukla2015introduction,chaubey2021positive,chaubey2022coulomb}) and the same mass $m_d$. These particles are confined vertically with a parabolic potential energy ${\mathcal{U}}^{ext}_z$ as well as horizontally with a ring-shaped potential energy ${\mathcal{U}}^{ext}_r$. Therefore, the Hamiltonian $\cal{H}$ of the system can be expressed as
 \begin{equation}
     \label{eq:Hamiltonian1}
{\cal{H}}={\cal{K}}_d+{{\mathcal{U}}^{ykw}}+{\mathcal{U}}^{ext}_z+{\mathcal{U}}^{ext}_r~{.}
\end{equation}
The Hamiltonian ${\cal{H}}$ is the sum of the kinetic energy ${\cal{K}}_d$, the interparticle interaction potential energy ${{\mathcal{U}}^{ykw}}$, and external potential energies ${\mathcal{U}}^{ext}$.
\subsection{Yukawa interparticle interaction potential}
\label{subsec:}
The dust interparticle interaction is govern by a Yukawa potential of the form
\begin{equation}\label{eq:pot_yuk}
{U^{ykw}_{ij}={\frac{Q}{r_{ij}}}exp(-{r_{ij}}/{\lambda_D})}{~.}
\end{equation}
 Here, $r_{ij}$ is the radial distance between two particles and ${\lambda_D}$ is the Debye length due to the background plasma \cite{hamaguchi1997triple}. Such Yukawa system can be thermodynamically described by two dimensionless parameters: the screening parameter ${\kappa}={a}/{\lambda_D}$ (i.e, the ratio of the inter-particle distance over the Debye length) and the unscreened Coulomb coupling parameter $\Gamma={Q^2}/{a{k_B}{T_d}}$ (the ratio of interparticle Coulomb energy to the thermal kinetic energy), here the interparticle distance is given by $a=({3}/{4{\pi}{n_d}})^{1/3}$, $n_d$ is the dust density, $T_d$ is the dust temperature, and $k_B$ is the Boltzmann constant.

\subsection{Vertical confinement from a parabolic potential well}
\label{subsec:}

\begin{figure}
\center	
\includegraphics[width=1.0\linewidth]{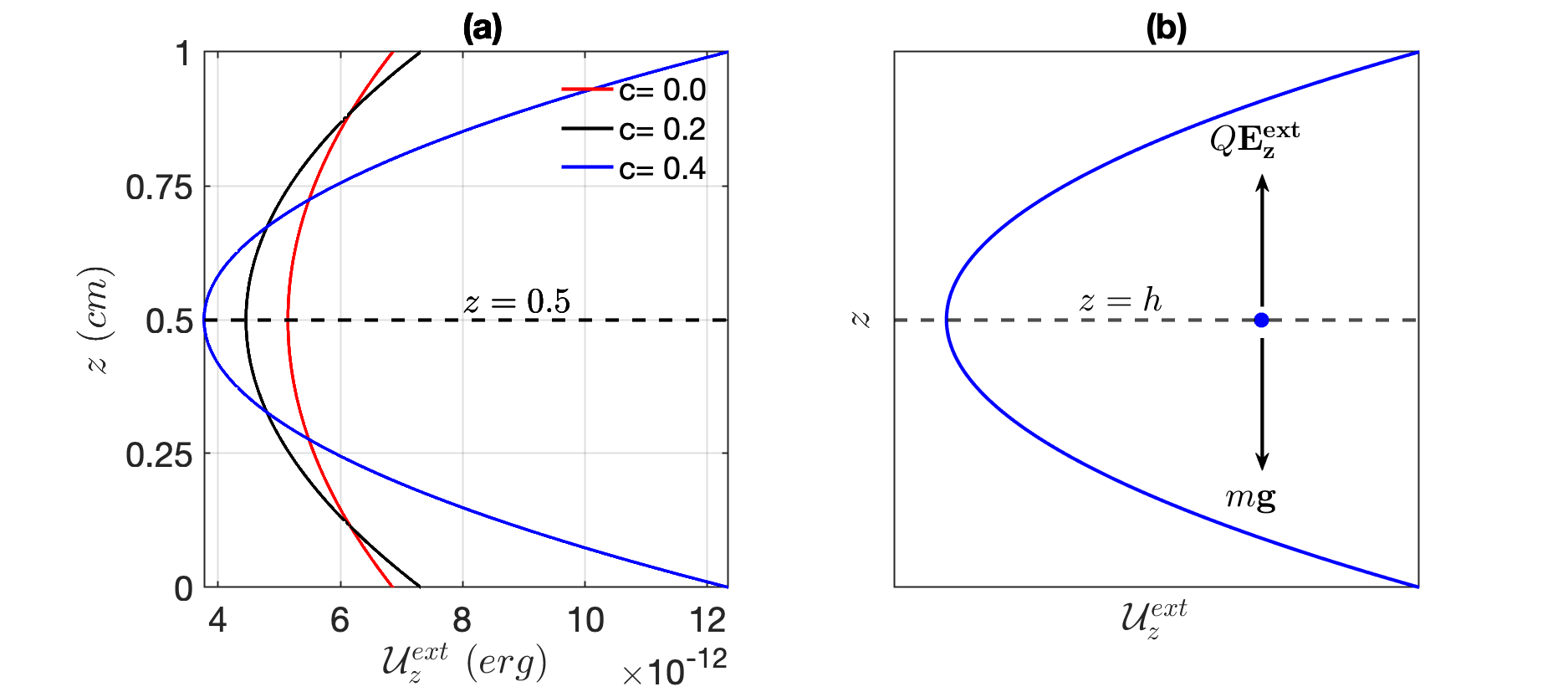}
    \caption{The net parabolic potential energy diagram. Subplot (a) shows the potential energy ${\cal{U}}^{ext}_z$ as a function of $z$ for several typical values of c and an equilibrium position $h=0.5~cm$. Subplot (b) shows a schematic diagram for an equilibrium position $z=h$ (horizontal dotted line) for a particle (blue dot) where the gravitational force is balanced by the force due to the parabolic potential energy (blue solid curve).}
	\label{fig:fig1}
\end{figure}

The simulation extends from 0 to $lz$ along the vertical $\hat{z}$ direction.  The external force due to gravity acts vertically downward 

\begin{equation}
   {{\bf{F}}^g_{z}}={{m_d}g{(-\hat{z})}} \nonumber
\end{equation}
 while the vertical upward electric force $F^{ext}_z$ is given by
\begin{equation}\label{eq:ext_vert_force}
F^{ext}_z={Q}E^{ext}_z {(\hat{z})}{.} \nonumber
\end{equation}
 Here we consider an electric field of the form
\begin{equation}\label{eq:ext_vert_efld1}
   E^{ext}_z={E_{z0}}{(z-lz+c)}{,} 
\end{equation}
  where $c<{{lz}/2}$ is a parameter that controls the curvature (“sharpness” and “depth”) of potential energy well. The larger the value of $c$, the sharper and deeper the curved potential well (see Fig~\ref{fig:fig1}(a)). In order to levitate the dust particles at an equilibrium vertical position $z=h$, the above two external forces should balance each other ~\cite{sukhinin2008non} at $z=h$~$i.e.$
\begin{equation}
    {{F}^g_{z}}=F^{ext}_z~{\text{at}}~{{z=h}}{.} \nonumber
\end{equation}
 Thus, the magnitude of the electric field at $z=h$ is given by
\begin{equation}\label{eq:ext_vert_efld2}
{E_{z0}}={\frac{{m_d}g}{Q}}{\frac{1}{(h-lz+c)}}
\end{equation}
 The total potential energy associated with each particle  at any vertical position $z$ is
\begin{equation}\label{eq:ext_vert_pot_eng1}
    {\mathcal{U}}^{ext}_z={{m_d}gz}+{Q}{U^{ext}_z}
\end{equation}
 and has contributions from the gravitational potential energy ${m_d}gz$ and the electrostatic energy ${Q}{U^{ext}_z}$ associated with the externally applied electric field $E^{ext}_z$. Here, $ {U^{ext}_z}=-{\int}{E^{ext}_z}{dz}$. 
 Therefore, the total potential energy becomes
\begin{equation}\label{eq:ext_vert_pot_eng2}
    {\cal{U}}^{ext}_z={{m_d}gz}-{{E_{z0}}}{\frac{(z-lz+c)^2}{2}} 
\end{equation}
This total potential energy has a parabolic shape with a symmetry axis perpendicular to gravity and centered on the equilibrium vertical position $z = h$. This parabolic potential is appropriate for modeling dust particles levitated near the plasma sheath in laboratory experiments ~\cite{qiao2005structural,melzer2006zigzag,hyde2013helical}. The form of the potential selected here has been used to determine analytically the dust particle charge \cite{tomme2000parabolic} and to simulate numerically structural transitions~\cite{totsuji1997structure_prl,totsuji1997structure_jap,maity2019molecular}. The potential energy diagram ($z$ vs ${\cal{U}}^{ext}_z$ ) for several values of $c$ is shown in Fig~\ref{fig:fig1}(a), where $h=0.5$ while the other parameters are fixed. This shows that as $c$ increases, the walls of the potential well become steeper and the well becomes deeper without changing the vertical location of the minima $h=0.5$. Thus, it is expected that a larger value of $c$ will result in a decreased volume of vertical space available to levitate the dust particles. Later in the paper, we show how an increase in the $c$ value can be used to simulate the formation of additional rings in the cylindrical ring structure. A schematic diagram in Fig~\ref{fig:fig1}(b) shows the equilibrium position for a particle (blue dot)  along the  horizontal dotted line at $z=h$ where the gravitational force is balanced by the confinement force due to the parabolic potential energy (blue solid curve).

\subsection{Horizontal confinement from a ring-shaped potential well}
\label{subsec:}

\begin{figure}
\center
    \includegraphics[width=1.0\linewidth]{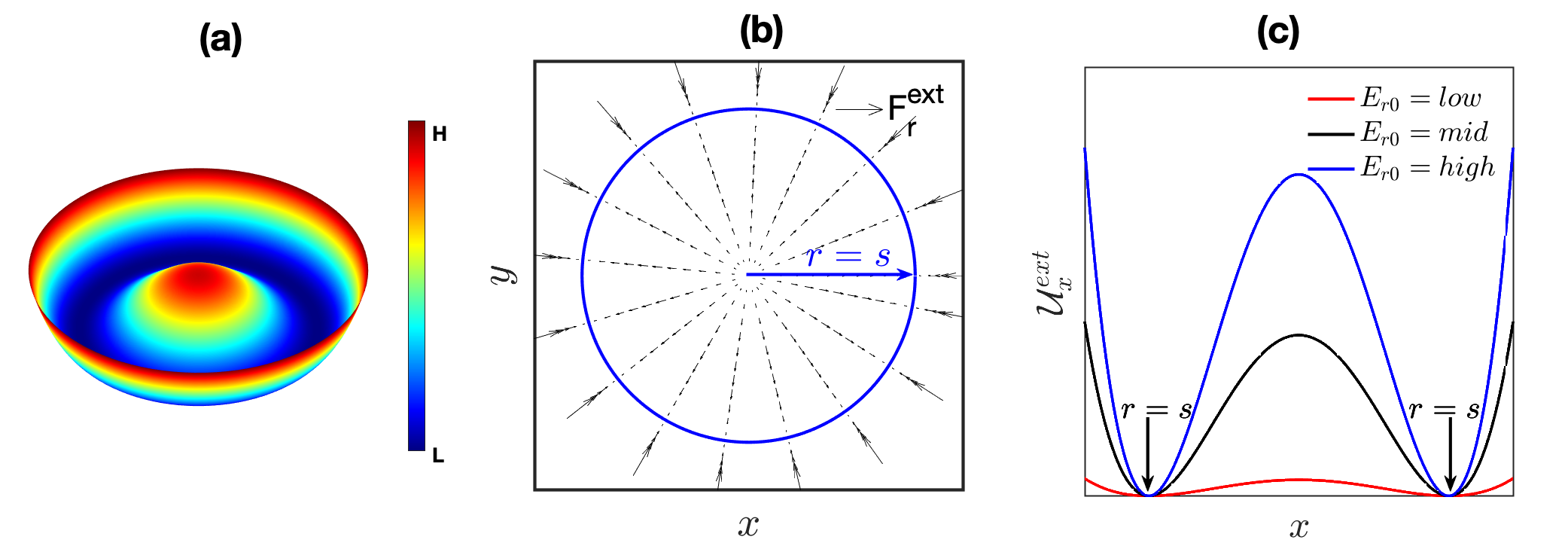}
	\caption{Radial force and potential diagrams (not to scale). (a) A $3d$ schematic surface diagram of the quatric ring-shaped potential energy discussed in Eq.~\ref{eq:ext_pot}. The letter H that appears in the color bar is the acronym for the high-potential value region, while L stands for the low-potential region. (b) A schematic quiver plot of the radial force vector from ~Eq.(\ref{eq:ext_force}). (c) A $1d$ projection of the $3d$ potential shown in (a) for three different values of ${E_{r0}}$.}
	\label{fig:fig2}
\end{figure}

In this work, we are interested in modeling experiments where a radial confinement potential traps the dust particle in a circular groove. For this we chose the following electric field profile
\begin{equation}\label{eq:ext_efld}
 {\bf{E}}^{ext}_r={E_{r0}}({x{\hat{x}}+y{\hat{y}}})(r-s)(r+s) {,}
\end{equation}
 where the horizontal plane is defined by the $x$-axis and the $y$-axis. Here, ${E_{r0}}$ is the constant value of the radial electric field. The corresponding potential is
\begin{equation}\label{eq:ext_pot}
    U^{ext}_r=-\frac{{E_{r0}}}{4}{(r-s)^2}{(r+s)^2} \nonumber{.}
\end{equation}
 Therefore, the total radial potential energy associated with each negatively charged dust particles is 
\begin{equation}\label{eq:ext_pot_eng}
    {\cal{U}}^{ext}_r=Q{U^{ext}_r}={E_{r0}}\frac{{Q}}{4}{(r-s)^2}{(r+s)^2}{.}
\end{equation}
 This quartic potential energy well is characterized by a ring of minimum energy at a radial distance $r=s$, a barrier centered at $r = 0$, and an axis of symmetry parallel to the gravitational attraction. Figure~\ref{fig:fig2} (a) shows a three dimensional ($3d$) view of this potential, while Fig.~\ref{fig:fig2}(c) shows a one dimensional ($1d$) projection. The corresponding force on a negatively charged dust particle is given by ${\bf{F}}^{ext}_r={Q}{\bf{E}}^{ext}_r$
\begin{equation}\label{eq:ext_force}
    {{\bf{F}}^{ext}_r}={Q}{E_{r0}}({x{\hat{x}}+y{\hat{y}}})(r-s)(r+s) {.}
\end{equation}

 ${{\bf{F}}^{ext}_r}$ is a radial vector force and has no tendency to swirl (${\nabla}{\times}{{\bf{F}}^{ext}_r}=0$). A schematic quiver plot of this radial force (Fig.~\ref{fig:fig2}(b)) shows that the force (as arrows) is pointing in the direction of minimum radial potential energy, $i. e., $~$r=s$ (the blue solid circle). The arrowheads pointing outwards (inside the circle) from the center represent the radially outward push on particles due to the central barrier, while the arrowheads pointing inwards (outside the circle) represent the inward push due to the periphery of the potential. Thus, it is expected that, in this potential, the particles get trapped in a circular groove centered at radius $r=s$.

 To illustrate how an increase in the quartic potential amplitude results in a ring structural transition, we have sketched the $1d$ profile (along $x$, at $y=0$) of this potential in Fig.~\ref{fig:fig2}(c) for three different values of ${E_{r0}}$. It is evident that an increase in ${E_{r0}}$ leads to steeper sides of the the well and a decreased radial space where the particles can levitate at a particular height $z=h$. Furthermore, Eq.~\ref{eq:ext_force} shows that higher ${E_{r0}}$ results in greater radial force, which tends to squeeze the particles closer to the minima of the well. Starting with a small amplitude, ${E_{r0}=low}$, and fixed parameters for the vertical potential, we can introduce sufficient number of dust particles to fill up the potential well and form a ring-shaped monolayer (co-planar nested rings of decreasing radius), radially centered around $r=s$ and vertically located at some $z=h$. The dust particles remain in-plain on the condition that, for the equilibrium interparticle spacing within the monolayer, the dust-dust interaction forces are balanced by the radial potential force and the gravitational force is balanced by the vertical parabolic potetnial force. As the amplitude of the radial confinement is increased to some intermediate value ${E_{r0}=mid}$, the dominance of the radial potential force over the dust-dust interactions will result in a structural state where these co-planner rings will show irregularities in the horizontal and vertical directions ($i.e$, transitional state). Finally, beyond a critical value of the amplitude, ${E_{r0}=high}$, new rings with the same diameter will form and align one above the other within a cylindrical surface with radius $s$ ($i.e$, a cylindrical shell structure). The new force balance will result in a new interparticle separation within the cylindrical shell.

\subsection{Governing equation}
\label{subsec:}
Including the interaction and confinement forces discussed above, the equation of motion of the $i^{th}$ dust particle for the Hamiltonian Eq.~(\ref{eq:Hamiltonian1}) can be written as 
 \begin{equation}
     \label{eq:langevin_equ}
	{m_d}{\ddot{\bf{r}}_i}={{\bf{F}}^{ykw}_{ij}}+
   {{\bf{F}}^{ext}_{r}}+{{\bf{F}}^g_{z}}+ {{\bf{F}}^{ext}_{z}}{.}
\end{equation}
 The right-hand side (RHS) of Eq.~(\ref{eq:langevin_equ}) is the sum of all the forces acting on the $i^{th}$ dust particle, which are given by
 \begin{eqnarray}\label{eq:langevin_force}
	{{\bf{F}}^{ykw}_{ij}}
	&=&-{\nabla}{\sum\limits_{\substack{i<j}}{{{U}}^{ykw}_{ij}}} \nonumber\\ 
	{{\bf{F}}^{ext}_{r}} &=&+{Q}{E^{ext}_r}{\hat{r}} \nonumber\\
 	{{\bf{F}}^g_{z}} 
 	&=&-{{m_d}g{\hat{z}}} \nonumber\\
   {{\bf{F}}^{ext}_{z}} 
    &=&+{Q}{E^{ext}_z}{\hat{z}}  \nonumber
\end{eqnarray}

 ${{\bf{F}}^{ykw}_{ij}}$ is the dust-dust interaction force, which is assumed to be Yukawa (screened Coulomb). The following three force terms account for radial confinement, gravity, and vertical confinement. In each simulation case presented here, we first disperse $N$ identical dust particles randomly in a $3d$ simulation box. Then, the particle dynamics is advanced according to  Eq.~(\ref{eq:langevin_equ}).

Here, we consider two cases: (i) ring structural transition due to changing amplitude of the quartic potential and (ii) properties of the cylindrical shell structure for various particle number, coupling, and curvature of the parabolic potential. The equilibrium state in each simulation run has been achieved using a Nose-Hoover thermostat \cite{nose1984molecular,hoover1985canonical}. The velocities were chosen to follow a Gaussian distribution corresponding to dust temperature $T_d$ for the considered coupling parameter $\Gamma$. It should be noted that for each simulation run, the system reached the desired equilibrium temperature (verified by temperature fluctuations and energy plots) well in advance of the simulation time.

First, to study the ring structural transition, we fix all other parameters and gradually change the amplitude ${E_ {r0}}$ of the quatric potential. This is achieved by a succession of simulation runs. For each value of ${E_ {r0}}$, the simulation is advanced until an equilibrium state is achieved. Then, the simulation is stopped, the potential amplitude is varied, and the simulation is advanced until a new equilibrium state is achieved. In these successive runs, the equilibrium particle positions at the last time step of one run are used as initial positions for next run (with a new ${E_ {r0}}$). The initial temperature remains the same for all simulation runs. The total sum of potential energy and kinetic energy is conserved in each individual run but varies from run to run due to the changing confinement potential energy.  

Next, for fixed quartic potential parameters, we explore how the properties of the cylindrical shell structure change with particle number, coupling strength, and characteristics of the parabolic (vertical) potential well.  First, we select a value of the quartic potential amplitude ${E_ {r0}}$ that results in the formation of a cylindrical shell structure for given initial particle number, coupling, and parabolic potential. After equilibrium is achieved, successive simulation runs are used to vary one parameter (particle number, coupling strength, or parabolic potential curvature), while all other parameters are fixed. In each case, the resulting formation of perfect or imperfect rings inside this shell structure is analyzed.

\FloatBarrier
\section{Numerical simulation, Results, Discussion}
\label{Num_Result_Dis}

 All simulations have been carried out using the open-source MD code LAMMPS \cite{plimpton1995fast}. Boundary conditions are periodic in the $x-y$ plane and non-periodic in the $z$ direction. For all simulations,  we consider a $3d$ simulation box of $lx~(=1cm){\times}ly ~(=1cm){\times}lz~(=1cm)$. $lx$, $ly$ and $lz$ are the system lengths in the $x$, $y$  and $z$ directions, respectively. Here, $-0.5{\le}~x~{\le}0.5$, $-0.5{\le}~y~{\le}0.5$ and $0{\le}~z~{\le}1$.  Each dust particle has the same charge $Q =11940~e^{-}$ and same mass $m_d=6.99{\times}10^{-10}g$. The confining quartic potential has a ring-shaped minimum along the radial distance $s=0.3535~cm$, located between the central barrier at $r=0~cm$ and the radial edge at $r=0.5~cm$. We have considered a constant value of gravitational acceleration, $g=981~cm/sec^2$. The other parameters are chosen according to the specific problem. \par
 
\subsection{Ring structure transition}
\label{subsec:}
 
  Keeping in mind the cylindrical symmetry of the quartic potential, at $t=0$, $96~(=3{\times}32)$ dust particles are randomly dispersed in a cylindrical volume of radius $r=0.5~cm$ centered around $(0,0)$ and extending on the $z$ axis from $z =0.4~cm$ to $z =0.6~cm$ within the simulation box. The coupling parameter is $\Gamma=200$ and the respective dust kinetic temperature $T_d={Q^2}/{a{k_B}{\Gamma}}=1.404~eV$. With typical inter-dust distance $a=7.31{\times}10^{-2}cm$ and ${\kappa}={a}/{\lambda_D}=1.0$, the corresponding value of the Debye length is ${\lambda_D} = 7.31{\times}10^{-2}cm$. The characteristic frequency of the dust particles is $\omega_{pd}={4{\pi}{n_d}Q^2}/{m_d}{\sim}19.0~sec^{-1}$ , which corresponds to the dust plasma period of $0.331~sec$ ($=2\pi/{\omega_{pd}}$). We have chosen a simulation time step of ${\Delta{t}}=5{\times}10^{-3}~{\omega}_{pd}^{-1}$ or ${\Delta{t}}=2.63{\times}10^{-4}~sec$ so that the simulation can easily resolve all phenomena occurring at the dust response time scale.

These initial conditions were used to study the ring structural transition due to a gradual change in the amplitude of the quartic potential. For each value of the quartic potential, the simulation is run until an equilibrium state is achieved. Each simulation run has a total number of time steps $N_{steps}=6{\times}10^{4}$, which corresponds to run time $15.78~sec$~($i.e.$, $N_{steps}{\times}{\Delta{t}}~sec$). After $t=15.78~sec$, the simulation is stopped and the amplitude of the quartic potential $E_{r0}$ from Eq.~\ref{eq:ext_pot} is changed. First, the amplitude is increased in $80$ successive runs from $E_{r0}=0.05~{statV/cm}$ to a maximum value of $E_{r0}=4~{statV/cm}$ (=$0.05{\times}80$) at $t=1262.4~sec$. Then, for the next $79$ runs, the amplitude is decreased by the same amount at the same time intervals to get back to $E_{r0}=0.05~{statV/cm}$. In this section, the curvature of the parabolic potential is kept fixed at $c=-1.0$. 
 
\begin{figure}
\center
	\includegraphics[width=1.0\linewidth]{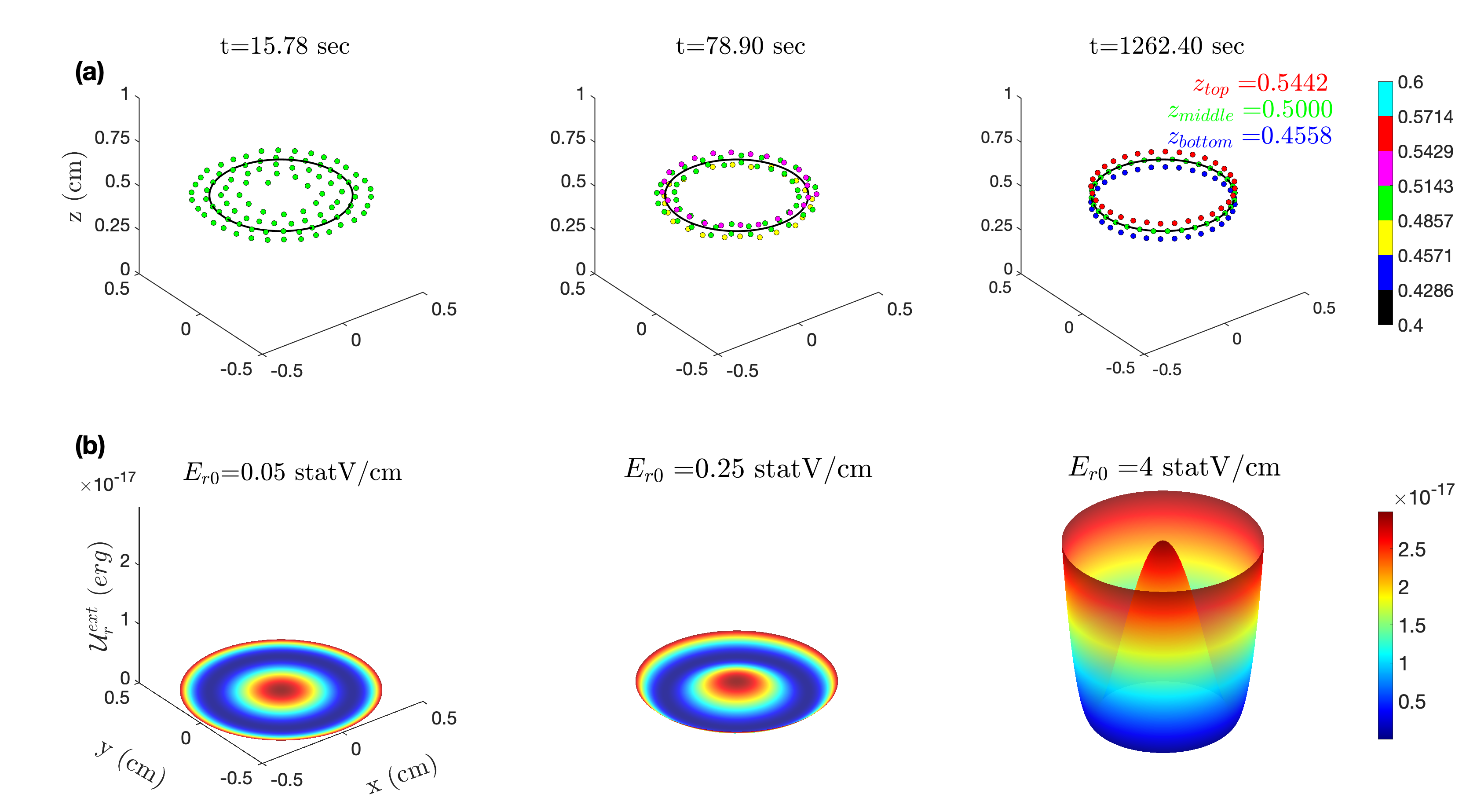}
	\caption{Three equilibrium stages of the ring structural transition are represented in (a) under the influence of the increasing amplitude of the ring-shaped quatric potential shown in (b).}
	\label{fig:fig3}
\end{figure}

 The three subplots in figure~\ref{fig:fig3}(a), from left to right,  display the final thermal equilibrium of $96$ dust particles for three representative values of the quartic potential well amplitude. The subplots in Fig.~\ref{fig:fig3}(b) show the shape of the corresponding potential well ${\cal{U}}^{ext}_r$ for each case. In Fig.~\ref{fig:fig3}(a), the locations of the quartic potential minima are marked by a black circles with radius $s=0.3535~cm$ at a vertical height $z=0.5~cm$ (minima of the parabolic potential). The colorbar in Fig.~\ref{fig:fig3}(a) corresponds to the vertical positions of the dust particles. In Fig.~\ref{fig:fig3}(b), the colorbar represents the magnitude of the quartic potential energy. During the first $15.78~secs$ of the simulation, the dust particles form a {\it{circular monolayer structure}} under the action of the quartic potential with amplitude $E_{r0}=0.05~{statV/cm}$. The monolayer consists of approximately four rings, nested within the same plane, which is visible by the same green color of the dots in the first subplot of Fig.~\ref{fig:fig3}(a). This circular structure has approximate symmetry about the $s=0.3535~cm$ and is located at a vertical height $z=0.5~cm$ because the net quartic potential force points in the direction of the minimum (as discussed in Sec.~\ref{model_methods}). The dust particles will remain within the plane if their planar interparticle separation is sufficient for the dust-dust interaction force to balance the confinement force from the quartic potential.

 The particle positions of this circular monolayer structure have been used as the initial positions for the next simulation (from $t=15.78~sec$ to $t=31.56~sec$) during which the quartic potential amplitude is increased from $0.05~{statV/cm}$ to $E_{r0} = 0.1~{statV/cm}$. Higher $E_{r0}$ value causes stronger radial force which radially squeezes the dust particle structure around the potential well's minimum. The central barrier of the potential pushes the ring of the innermost particles towards the minimum, while the circular edge pushes the ring of the outermost particles towards the minimum. The competition between the radial potential (due to the increase in amplitude) and the dust-dust interactions results in irregularities in the co-planner rings both in the horizontal (perpendicular to gravity) and vertical (parallel to gravity) directions.

At the later time $t=78.90~sec$, as the net confinement force increases by increasing the amplitude to $E_{r0} = 0.25~{statV/cm}$, the middle two rings shift along the vertical z-axis, one towards the bottom (yellow dots), one towards the top (magenta dots). The innermost ring (which now has larger radius than earlier at $t=15.78~sec$) and outermost ring (which now has smaller radius than earlier at $t=15.78~sec$) remain nearly co-planar (green dots) at the same vertical height $z = 0.5~cm$. We refer to this arrangement of dust particles as a {\it{transitional structure state}}. An expanded image of the transitional structure state is shown in figure~\ref{fig:fig5}(b).

Further increasing the value of $E_{r0}$ eventually leads to a structure in which the dust particles are aligned in rings with same diameter, located one above the other within a cylindrical surface. At $t=1262.4~sec$, when the potential amplitude is  $E_{r0} = 4~{statV/cm}$, the particle settle in three perfect rings with a radius of $0.3535~cm$: the top one at $z=0.5442~cm$ (red dots), the middle one at $z=0.5~{cm}$ (green dots), and the bottom one at $z=0.4558~cm$ (blue dots). We refer to this state as a {\it{cylindrical shell structure}}, where each of the three rings has an equal number of dust particles (here, $32$.) Subsequent simulation runs with a higher  potential amplitude do not yield discernible changes to the cylindrical ring structure. All transitional states in between the circular monolayer structure and cylindrical shell structure consist of irregular rings that manifest zig-zag instabilities in both the horizontal and vertical directions. 

\begin{figure}
\center
	\includegraphics[width=1.0\linewidth]{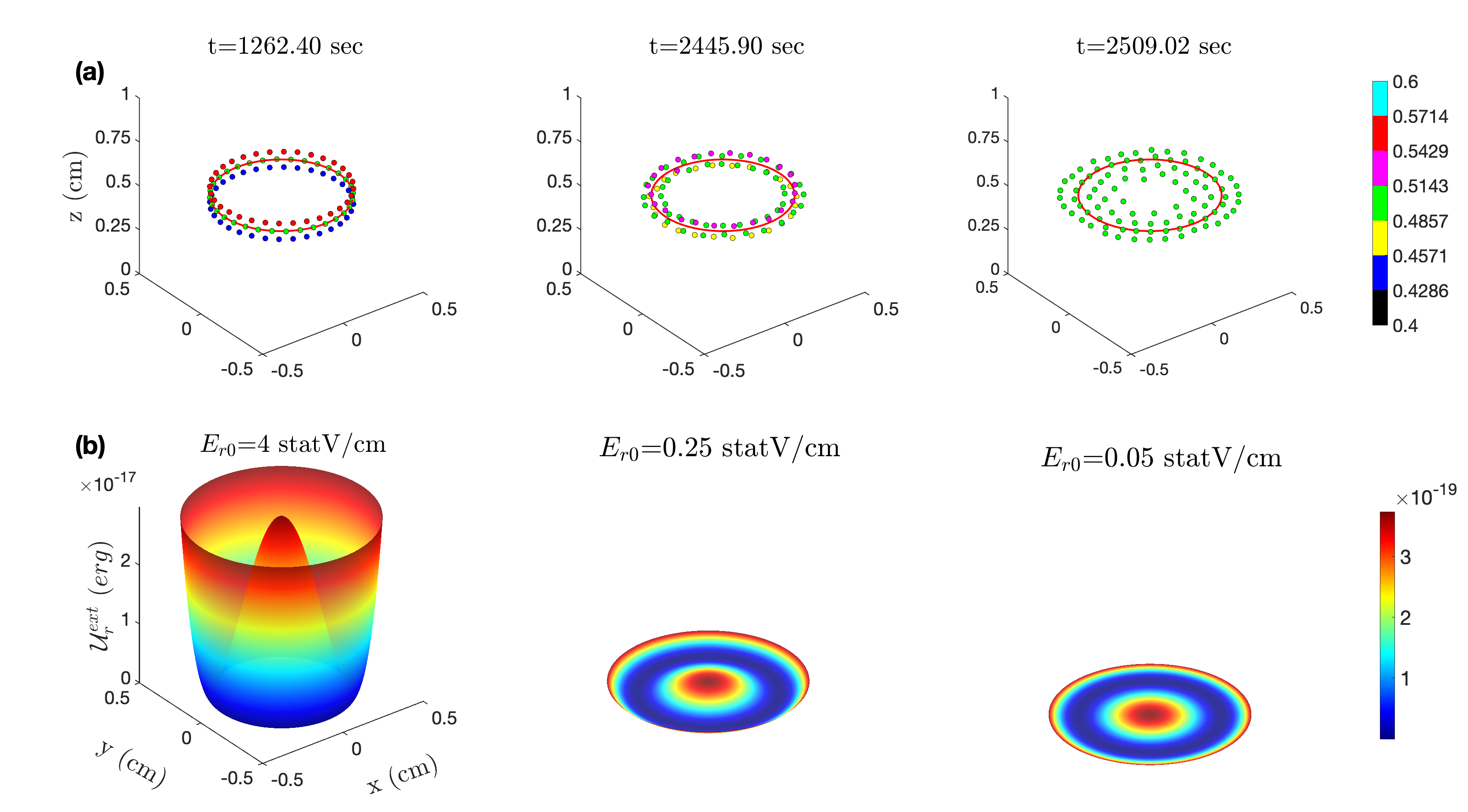}
	\caption{Three equilibrium stages of the ring structural transition are represented in subplot (a) under the influence of the decreasing amplitude of the ring-shaped quartic potential shown in subplot (b).}
	\label{fig:fig4}
\end{figure}

\begin{figure}
\center	
\includegraphics[width=1.0\linewidth]{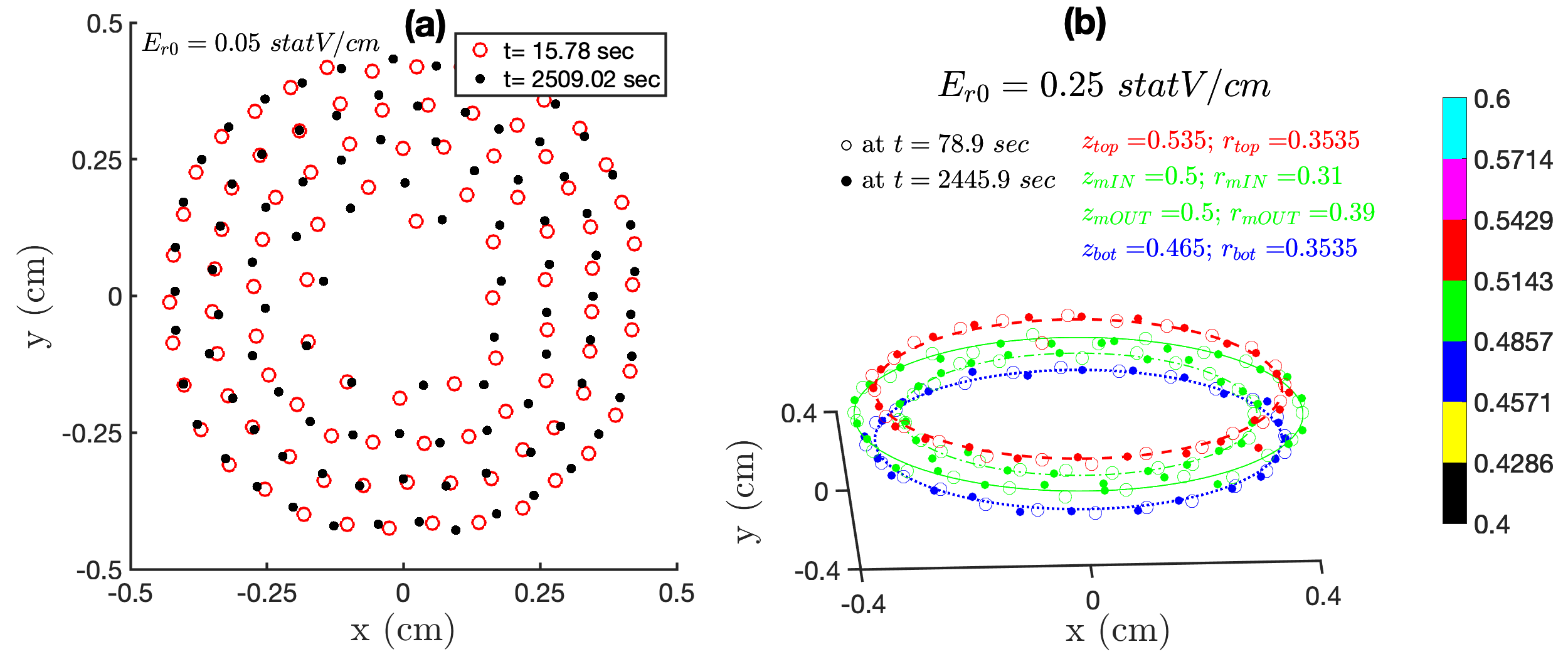}
	\caption{ Comparison of the dust position distributions from the forward and backward simulations for corresponding values of $E_{r0}$. The plot in (a) shows the distribution of particle positions for $t=15.78~sec$ (red circles, Ref. Fig.~\ref{fig:fig3}(a)) along with the distribution at $t=2509.02~sec$ (black dots, Ref. Fig.~\ref{fig:fig4}(a))). The plot in (b) shows the distribution of particle positions for $t=78.90~sec$ (multi-color circles, Ref. Fig.~\ref{fig:fig3}(a)) along with the distribution at $t=2445.90~sec$ (multi-color dots, Ref. Fig.~\ref{fig:fig4}(a)). Colorbar in the right panel corresponds to the vertical locations (z) of the particles.}
	\label{fig:fig5}
\end{figure}

To test the reversibility of this structure transition phenomenon, we repeat the simulation from $t=1262.4$ to $t=2509.02~sec$, this time decreasing the amplitude value from $E_{r0}=4~{statV/cm}$ to $E_{r0}=0.05~{statV/cm}$ by the same amount of $0.05~{statV/cm}$ at the same interval of time ($=15.78~sec$) for each simulation run. The outcomes of this simulation are shown in Fig.~\ref{fig:fig4}. Based on a visual comparison of subplots, the formation of structures in Fig.~\ref{fig:fig4}(a) appears to be similar to that in Fig.~\ref{fig:fig3}(a) for the same values of $E_{r0}$. Thus, we find that this structure transition is reversible in terms of layer or ring formation.

 As we mentioned above, in each successive run, the final particle positions of the previous state have been used as the initial positions for the next one. This means that during the backward and forward simulations, the initial particle positions were not the same for the same $E_{r0}$. For example, during the forward run for $E_{r0}=0.05~{statV/cm}$ (at $t=15.78~sec$ in Fig.~\ref{fig:fig3}(a)), the initial particle positions were random, while during the backward run, the initial particle positions for $E_{r0}=0.05~{statV/cm}$ (at $t=2445.9~sec$ in Fig.~\ref{fig:fig4}(a)) belong to the previous state for $E_{r0}=0.1~{statV/cm}$ (at $t=2493.42~sec$). Therefore, it is expected that the equilibrium particle distributions for forward and backward runs with the same $E_{r0}$ will differ.

 Fig.~\ref{fig:fig5} shows the equilibrium states from the forward and backward simulation runs with the same $E_{r0}$. The distribution of dust particles at $t=15.78~sec$ (Ref. Fig.~\ref{fig:fig3}(a)) from the forward simulation is represented by red circles (${\color{red}{\circ}}$), while the distribution of particles at $t=2509.02~sec$ (Ref. Fig.~\ref{fig:fig4}(a)) from the backward simulation is represented by black dots ({$\bullet$}) together in the same Fig.~\ref{fig:fig5}(a). We see that, although both simulations resulted in circular monolayer structures, the distribution of final particle positions is different. This is due to the differences in initial distributions of particle positions in each simulation. Similar trend is observed for transitional states. In Fig.~\ref{fig:fig5}(b), the distribution of dust particles from the forward simulation (represented by the multi-colored circles (${{\circ}}$)) at $t=78.9~sec$ (Ref. Fig.~\ref{fig:fig3}(a)) has been plotted  over the distribution of particles from the backward simulation (represented by multi-colored dots ({$\bullet$})) at time $t=2445.9~sec$ (Ref. Fig.~\ref{fig:fig4}(a)). It is again visible that particle locations are not exactly the same under the same external potential $E_{r0}=0.25~{statV/cm}$. In Fig.~\ref{fig:fig5}(b), we have highlighted the four rings with solid circles based on their radii and vertical positions: the bottom (blue; radius $r_{bot}=0.3535~cm$ and vertical location $z_{bot}=0.465~cm$), the top (red; radius $r_{top}=0.3535~cm$ and vertical location $z_{top}=0.535~cm$), and the two middle rings (green; radii of the inner ($r_{mIN}=0.31~cm$) and outer $r_{mOUT}=0.39~cm$, and same vertical location $z_{mIN}=z_{mOUT}=0.5~cm$). Thus, we conclude that the ring structural transition exhibits hysteresis in the particle positions.

\begin{figure}
\center
\includegraphics[width=1.0\linewidth]{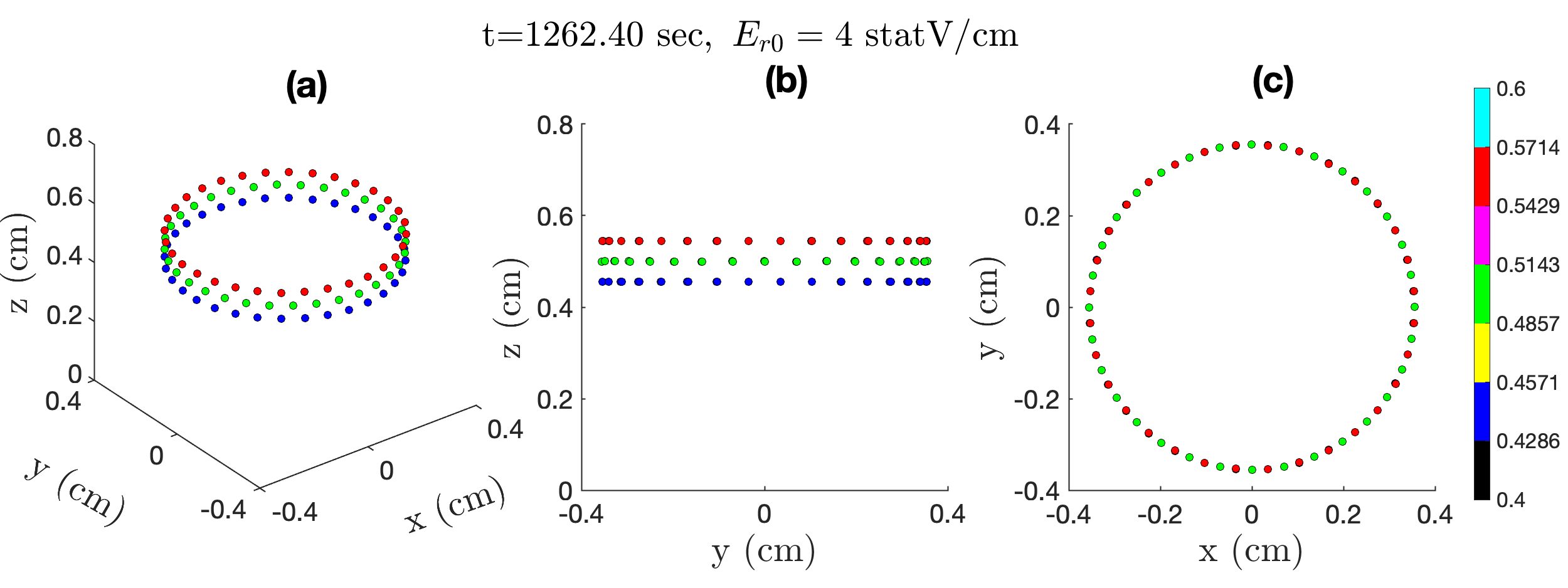}
\caption{The dust particles within the cylindrical surface structure arrange in a regular hexagonal pattern for $E_{r0}=4.0~{statV/cm}$ (data from $t=1262.40~sec$). The figure panels (Ref. Figs.~\ref{fig:fig3}(c) and ~\ref{fig:fig4}(a)) show three different views: (a) $3d$ view, (b) y-z plane, and (c) x-y plane.}
	\label{fig:fig6}
\end{figure}

 Another interesting observation is that within the cylindrical shell structure, the dust particles arrange in a hexagonal lattice structure, as shown in Fig.~\ref{fig:fig6}.  Figures~\ref{fig:fig6}(a) ($3d$ view) and \ref{fig:fig6}(b) ($y-z$ plane). The z-position of the top ring particles (red dots) aligns with the z-position of the bottom ring particles (blue dots), while the middle ring particles (green dots) are located in between. If the number of particles in the simulation is increased enough so that more rings are added to the cylindrical shell structure, the symmetry in vertical alignment will repeat, forming a A-B-A-B-A-B structure (the curvature of the parabolic potential (causes for the vertical confinement) may also affect this structure).

\begin{figure}
\center
\includegraphics[width=1.0\linewidth]{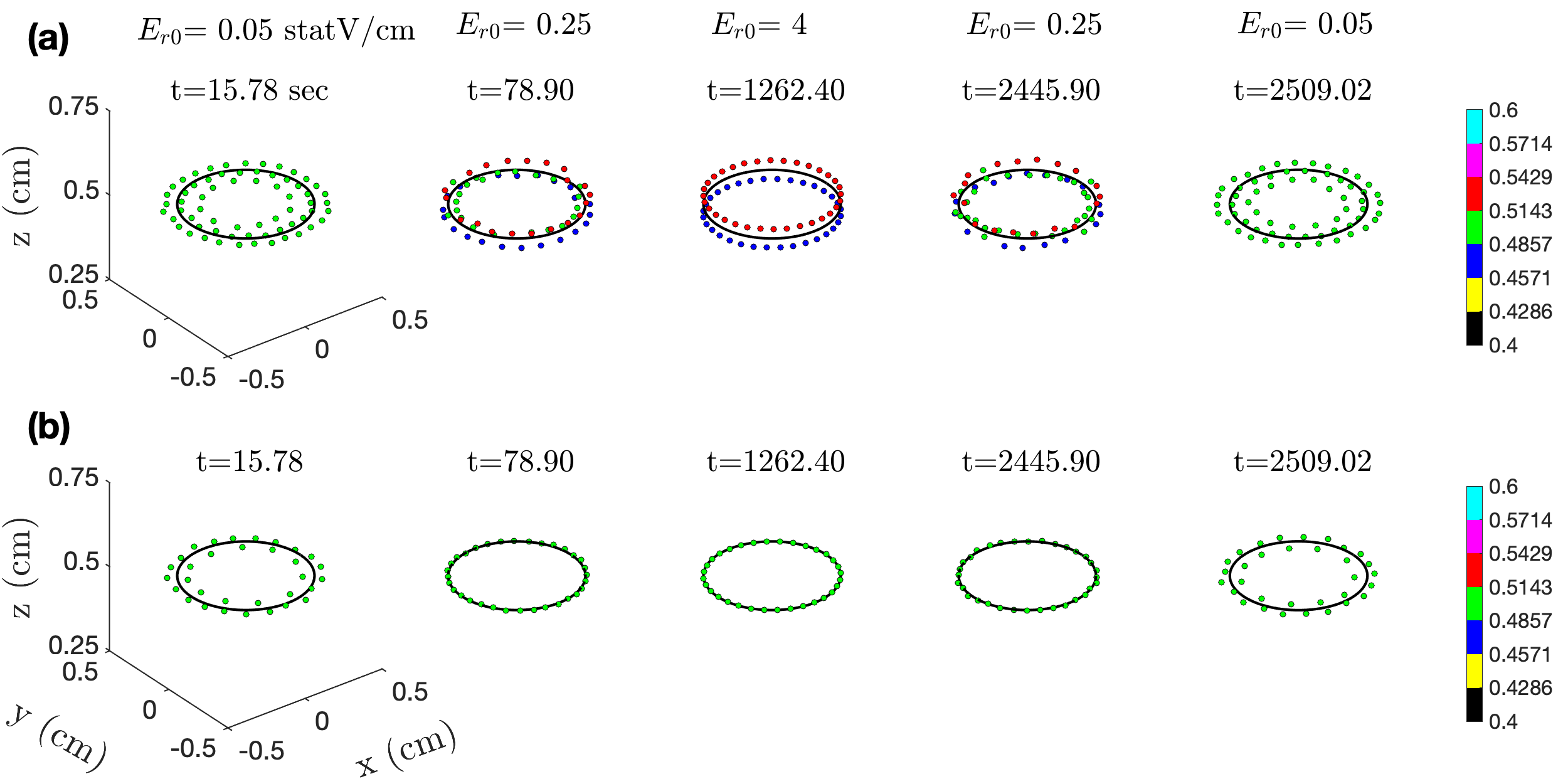}
	\caption{The ring structural transition for (a) $64~(=2{\times}32)$ and (b) $32~(=1{\times}32)$  dust particles due to the change in $E_{r0}$.}
	\label{fig:fig7}
\end{figure}

Figure~\ref{fig:fig7} shows the same simulation experiment repeated with smaller number of particles. Figure~\ref{fig:fig7}(a) depicts a ring structural transition for $64~(=2{\times}32)$ dust particles. The cylindrical shell structure in this case is made up of two perfect rings, each with 32 particles, where the top ring particles (red dots) settle in between the bottom ring particles (blue dots). Figure~\ref{fig:fig7}(b) depicts a $32~(=1{\times}32)$ particle with a single perfect ring at the maximum value of the potential amplitude. 

  \FloatBarrier
  \FloatBarrier

\subsection{Parametric dependency in the formation of the cylindrical shell structure}
\label{subsec:}

\begin{figure}
\center
\includegraphics[width=1.0\linewidth]{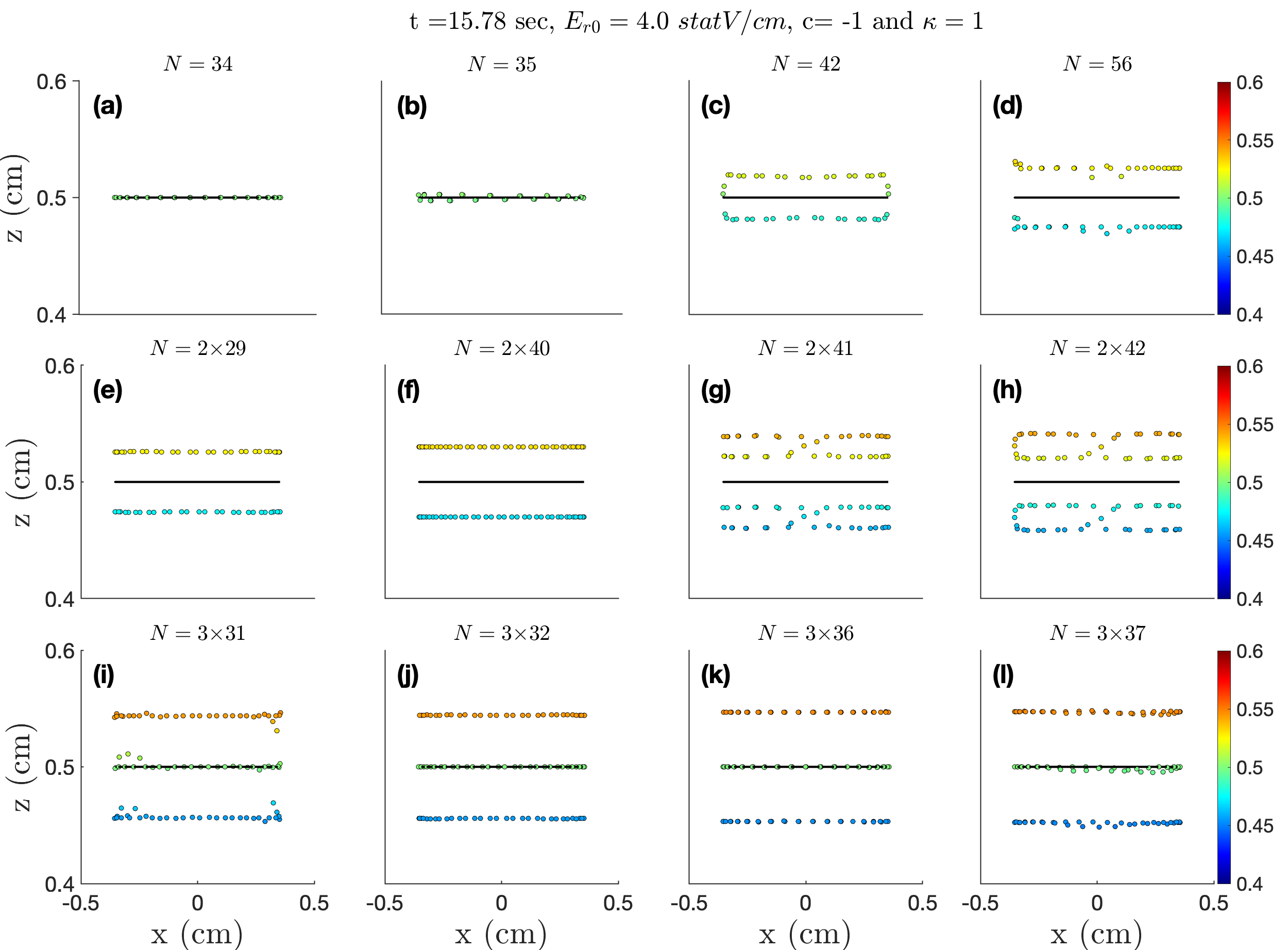}
	\caption{Ring structure formation in the cylindrical surface state for different number of dust particles $N$. All subplots show the equilibrium structure at $t=15.78~sec$ with fixed values of $E_{r0}=4.0~{statV/cm}$, $\kappa$=1, and $c=-1$.}
	\label{fig:fig8}
\end{figure}

\begin{figure}
\center
\includegraphics[width=1.0\linewidth]{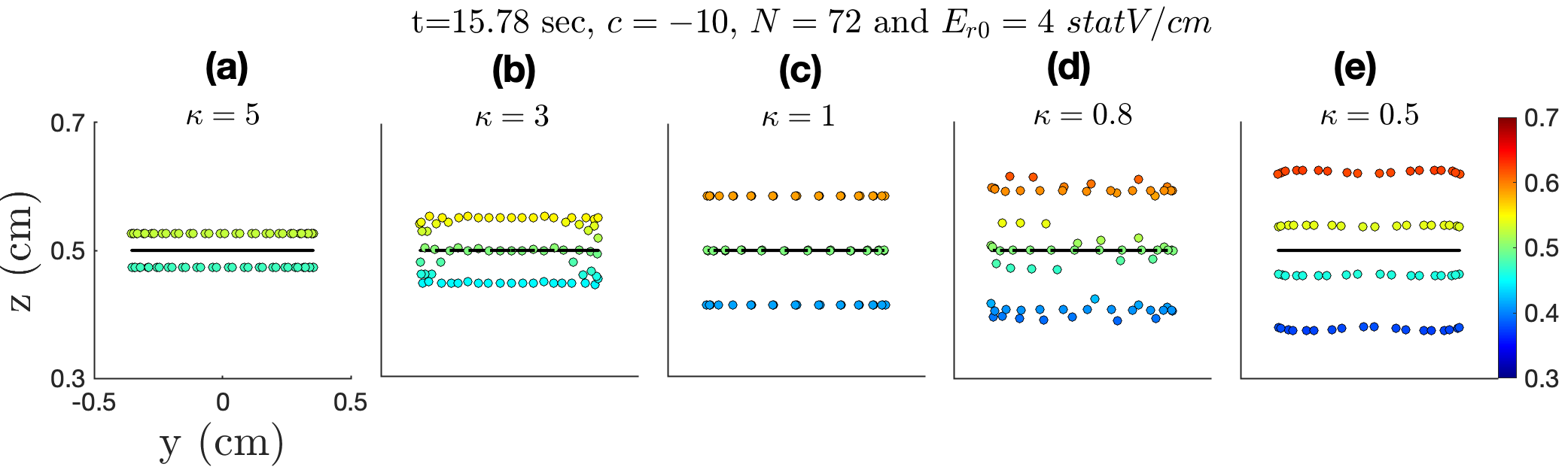}
	\caption{Ring structure formation in the cylindrical surface state for different $\kappa$. 
 All subplots show the equilibrium structure at $t=15.78~sec$ with fixed values of $E_{r0}=4.0~{statV/cm}$, number of particles $N=72$, and $c=-10$.}
	\label{fig:fig9}
\end{figure}

\begin{figure}
\center
\includegraphics[width=1.0\linewidth]{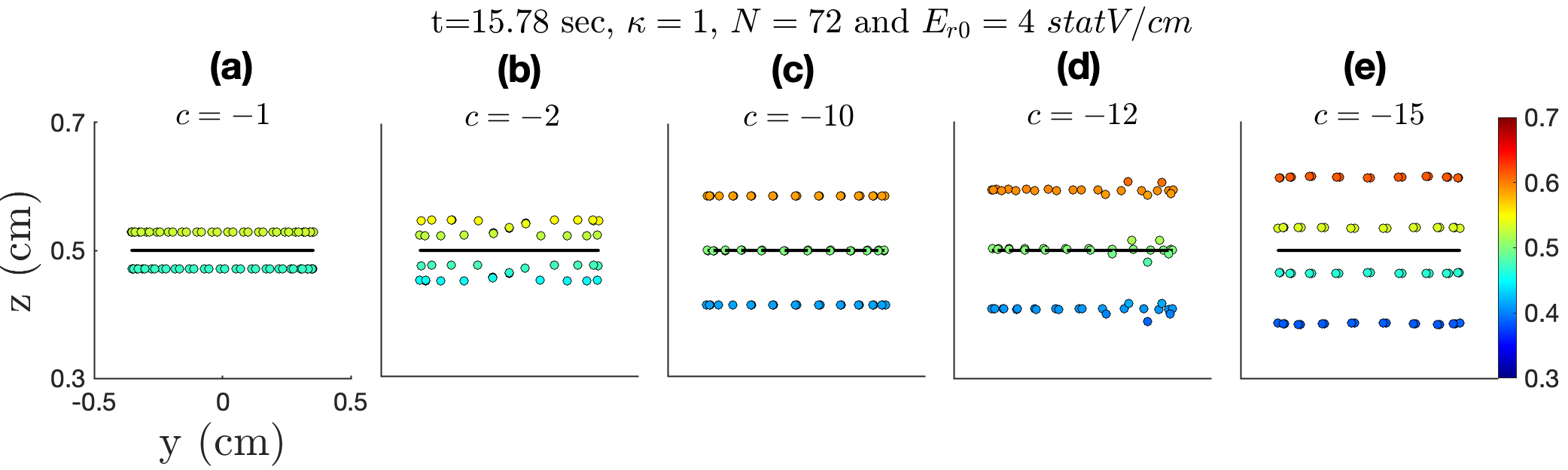}
	\caption{Ring structure formation in the cylindrical surface state for different $c$. All subplots show the equilibrium structure at $t=15.78~sec$ with fixed values of $E_{r0}=4.0~{statV/cm}$, number of particles $N=72$, and $\kappa=1$ are kept constant.}
	\label{fig:fig10}
\end{figure}

In addition to the amplitude value $E_{r0}$, the number of dust particles $N$ (or number density), the screening parameter $\kappa$, and the control parameter $c$ for parabolic potential ${\cal{U}}^{ext}_z$, all play a crucial role in the formation of the observed dust structures. To investigate the role of these parameters, all results in this section were obtained from a single simulation period ($0~sec$ to $15.78~sec$) with fixed amplitude $E_{r0}=4.0~{statV/cm}$. The initial state is a random distribution of dust particles which forms a cylindrical shell structure at equilibrium for the chosen amplitude $E_{r0}=4.0~{statV/cm}$. To ensure that the particles are distributed uniformly within each ring and to prevent a zigzag instability or ring irregularities in the cylindrical shell structure, the number of particles should be divisible by the number of rings. We have summarized some findings regarding the formation of perfect or imperfect rings in  the cylindrical shell structures as different parameters are varied: Fig.~\ref{fig:fig8} (changing $N$ for constant $\kappa=1$ and $c=-1$), Fig.~\ref{fig:fig9} (changing $\kappa$ for constant $c=-10$ and $N=72$) , and Fig.~\ref{fig:fig10} (changing $c$ for constant $\kappa=1$ and $N=72$) for a fixed value of $E_{r0}=4.0~{statV/cm}$.

 Fig.~\ref{fig:fig8} shows the results for different numbers of dust particles $N$ with fixed values of $\kappa$=1 and $c=-1$. As illustrated in Fig.~\ref{fig:fig8}(a), particle numbers up to $34$ arrange into a single perfect ring with radius $s=0.3535~cm$, located at vertical equilibrium height $z=0.5~cm$. As the density of the dust particles increases, the interparticle separation decreases and the repulsive Yukawa interaction becomes more pronounced. Since there is no more room to expand radially within the same plane, further increasing $N$ results in the extension of existing ring vertically with some irregularities and, eventually, in the formation of a new ring. Thus, the addition of one more particle to the $34$-particle ring the structure becomes irregular due to a zigzag instability, as displayed in Fig.~\ref{fig:fig8}(b). Further addition of particles leads to the formation of a new ring located at a different vertical height. Initially two irregular rings are formed that attempt to balance vertically: one shifts lower and the other shifts upper with respect to the vertical equilibrium location $z=0.5~cm$, as shown in Fig.~\ref{fig:fig8}(c)-(d). Finally, two perfect rings are observed for particle numbers ranging from $N=58=2{\times}29$ up to $N=80=2{\times}40$, as shown in Fig.~\ref{fig:fig8}(e)-(f).  The addition of more particles results in the formation of an additional ring through similar intermediate irregular structures (Fig.~\ref{fig:fig8}(g)-(i)). The formation of three perfect rings is observed for particle numbers ranging from $N=96=3{\times}32$ to $N=108=3{\times}36$, as shown in Figs.~\ref{fig:fig8}(j) and \ref{fig:fig8}(k). The formation of additional rings via irregular intermediate structures is repeated if more particles are added to the current system (Fig.~\ref{fig:fig8}($l$)).

The snapshots in figure~\ref{fig:fig9} represent the equilibrium configurations of $72$ particles for different $\kappa$ at $t=15.78~sec$. The values of $E_{r0}=4~{statV/cm}$ and $c=-10$ are constant. Figure~\ref{fig:fig9}(a) shows that two perfect rings, made of $36$ particles each, form for $\kappa=5$. The parameter $\kappa$ is the ratio of average interparticle separation to the Debye screening length. Thus it quantifies the characteristic spacial scale at which the Yukawa interaction acts. As $\kappa$ is decreased, the range of repulsive interactions among dust particles increases. For $E_{r0}=4~{statV/cm}$ it is not favorable for the dust particles to spread in the radial direction. Therefore, the effect of decreasing $\kappa$ can either lead to expanding the dust interparticle separation vertically, which results in ring irregularities, or to producing a new ring at a different vertical height. Figure~\ref{fig:fig9}(b) shows the formation of irregularities in the existing two rings as they try to form a third ring for $\kappa=3$. Finally, three perfect rings are observed to form for  $\kappa=1$, as shown in figure~\ref{fig:fig9}(b), with each ring containing $24$ particles ($N=72=3{\times}24$). The formation of a forth ring through intermediate irregular states is repeated as the value of $\kappa$  is further decreased, as shown in Figure~\ref{fig:fig9}(d)-(e). It should be emphasized that in this case, the value of $c$ must favor the addition of a new ring to the given structure. If $c = -1$, we don't see this transition event.

Figure~\ref{fig:fig10} shows equilibrium configurations of $72$ particles for different $c$ values with fixed values of $E_{r0}=4~{statV/cm}$ and $\kappa=1$ at $t=15.78~sec$. For $c=-1$, two perfect rings are observed, as shown in figure~\ref{fig:fig10}(a), and each of the rings has an equal number of dust particles, $36$. The parameter $c$ controls the curvature of the parabolic potential whose axis of symmetry is perpendicular to the gravitational force and centered on the equilibrium vertical position $z=0.5$. When $c$ decreases, the curvature of the parabolic confining potential becomes less steep, increasing the amount of vertical space is available for the particles. This can either cause growing irregularities between neighboring rings or result in the formation of new rings. Figure~\ref{fig:fig10}(b) shows that for $c=-2$, the interparticle separation in the existing two rings  increasess in the vertical direction resulting in irregular structure. For $c=-10$, figure~\ref{fig:fig9}(c) depicts the formation of a third ring, with each ring containing $24$ particles ($N=72=3{\times}24$). If the value of $c$ is further decreased, additional rings are formed through intermediate irregular states, as shown in Figure~\ref{fig:fig10}(d)-(e).

Table~\ref{table:table1} provides a summary of results for simulations with different number of dust particles $N$ with fixed values of $\kappa=1$. The range of perfect ring formation has is shown by rows for $c=-5$ and $c=-10$.
\begin{table}[h]
\begin{tabular}{|l||*{2}{c|}}\hline
\backslashbox{No. of rings}{{\hspace{-1.0cm}}c}
&\makebox[5em]{$c=-5$}
&\makebox[5em]{$c=-10$}
\\\hline\hline
{1}& {up to 23}& {up to 19}\\\hline
{2}& {$32~(=2{\times}16)$ to $54~(=2{\times}27)$}& {$24~(=2{\times}12)$ to $44~(=2{\times}22)$}\\\hline
{3}& {$60~(=3{\times}20)$ to $78~(=3{\times}26)$} & {$51~(=3{\times}17)$ to $72~(=3{\times}24)$}\\\hline
{4}& {$96~(=4{\times}24)$} & {$84~(=4{\times}21)$ to $92~(=4{\times}23)$} \\\hline
{5}& {-}& {$125~(=5{\times}25)$ to $130~(=5{\times}26)$} \\
\hline
\end{tabular}
\caption{Ring structure formation in the cylindrical shell state for different numbers of dust
particles for $c=-5$ and $c=-10$.}
\label{table:table1}
\end{table}

\section{Conclusions and Outlook}
\label{Conclusions}
In dusty plasmas, the competition between interaction potential forces and external confinement forces can lead to the formation of interesting structural states and structural transitions. To explore the ring structural transition in a dusty plasma, we conducted MD simulations where charged dust particles are confined in a ring-shaped quartic potential well. Here we presented the results from two cases. In the first case, we examined how increasing value of the quartic potential amplitude can lead to the transition from a ring monolayer structure (rings of different diameters nested within the same plane) to a cylindrical shell structure (rings of similar diameter aligned in parallel planes). We established that the ring structure transition occurs through several transitional states where the rings exhibit irregularities and zig-zag instabilities. The transition is also reversible, but shows hysteresis in the initial and final distributions of particle positions. In the cylindrical surface structure, the particles are arranged in a perfect hexagonal pattern, with each ring containing equal number of particles. 

In the second case, we investigated how the ring formation within the cylindrical surface structure state depends on the number of dust particles $N$, the screening parameter $\kappa$, and the curvature of the parabolic potential (denoted by the parameter $c$) for a fixed value of $E_{r0}=4~{statV/cm}$. For this high value of the quartic potential amplitude it is not energetically favorable for the particles to move in the radial direction, the effect of the above parameters results either in the formation of ring irregularities due to increased particle separation in the vertical direction, or in the formation of new rings at different vertical height. Therefore, as $\kappa$ decreases, $c$ decreases, and/or the number density increases, we observe the formation of additional rings within the cylindrical shell through several intermediate states with irregularities solely in the vertical direction. 

We suggest that the represent configuration (duty plasma confined in a quartic potential well) can be used to explore various fundamental phenomena. For example, in a laboratory experiment where the quartic potential is achieve by a combination of rings on the lower electrode and/or weak magnetic field, rapid variations of the electrodes power can be used to cause implosion or explosion of the annular dust structure that would allow the exploration of various phenomena, including acoustic waves, two-stream instabilities, bump-on-tail instabilities, spatial variation in coupling strength, and much more. This has been previously shown for sculpted ultracold neutral plasmas in \cite{dharodi2020sculpted}.  Since the transition from a ring monolayer to a cylindrical shell structure is highly sensitive to the dust interaction potential, we further conjecture that it can be used to investigate the dust particle charging in laboratory settings.

\section{Acknowledgments}
This work is supported by NSF 1903450 and NSF OIA-2148653 (EPSCoR FTTP).


\appendix
\section{Supplemental Material}

In the following Fig.~\ref{fig:fig_suply}, we compute the $96$ random dust particle system with the change in trend of $E_{r0}$ which is reversed from the previously discussed case in Fig.3 and Fig.4. This means that the starting amplitude is the highest, $E_{r0}=4~{statV/cm}$, then decreases to the lowest, $E_{r0}=0.05~{statV/cm}$, and then begins to increase to $E_{r0}=4~{statV/cm}$. Apart from the reverse order of the amplitude of potential, all other conditions are similar.  In Fig.~\ref{fig:fig_suply}, as expected, all five subplots show the transformation from a cylindrical shell structure ($E_{r0}=4~{statV/cm}$) to a circular monolayer structure ($E_{r0}=0.05~{statV/cm}$) and again to a cylindrical shell structure ($E_{r0}=4~{statV/cm}$) through various transition states.

\begin{figure}[H]
\center
\includegraphics[width=1.0\linewidth]{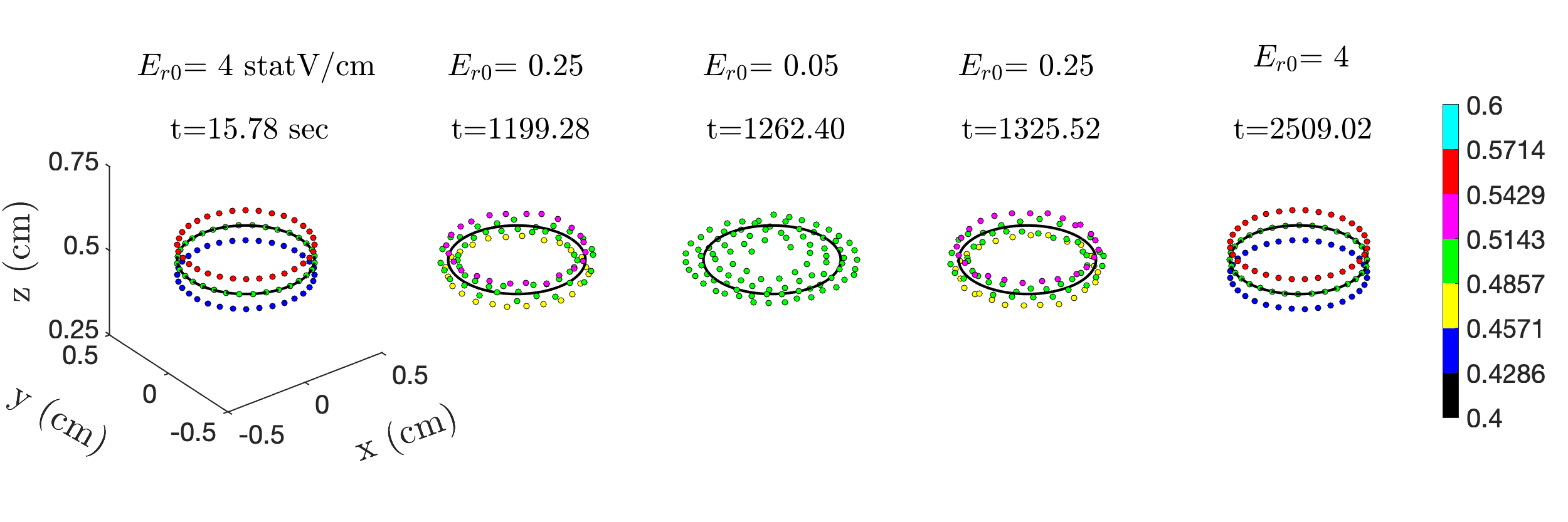}
	\caption{A ring structural transition phenomenon under containment swinging potential  well. There is an initial decrease in magnitude of quatric potential well and then an increase in the magnitude.}
	\label{fig:fig_suply}
\end{figure}

\end{document}